# Neutron Time-Of-Flight Spectrometer Based on HIRFL for Studies of Spallation Reactions Related to ADS Project

ZHANG Suyalatu[1,2], CHEN Zhiqiang[1,*], HAN Rui[1], WADA Roy[1], LIU Xingquan[1,2], LIN Weiping[1,2], LIU Jianli[1], SHI Fudong[1], REN Peipei[1,2], TIAN Guoyu[1,2], LUO Fei[1]

[1]Institute of Modern Physics, Chinese Academy of Sciences, Gansu, Lanzhou 730000, China

[2]University of Chinese Academy of Sciences, Beijing 100049, China

[*]Corresponding author. *E-mail address:* zqchen@impcas.ac.cn

**Abstract:** A Neutron Time-Of-Flight (NTOF) spectrometer based on Heavy Ion Research Facility in Lanzhou (HIRFL) is developed for studies of neutron production of proton induced spallation reactions related to the ADS project. After the presentation of comparisons between calculated spallation neutron production double-differential cross sections and the available experimental one, a detailed description of NTOF spectrometer is given. Test beam results show that the spectrometer works well and data analysis procedures are established. The comparisons of the test beam neutron spectra with those of GEANT4 simulations are presented.

**I.  Introduction**

A great interest of spallation reactions [1,2] has been increasing with the development of Accelerator Driven Systems (ADS) and the applications of Spallation Neutron Source (SNS). The spallation reaction is defined as interactions between a light projectile (e.g. proton) with the GeV range energy and heavy target nuclei which is split to a large number of hadrons (mostly neutrons) or fragments. Many nuclear models, such as intra-nuclear cascade-evaporation (INC/E) [3] model and quantum molecular dynamics (QMD) model [4], have been developed to study spallation reactions. These nuclear models are coded for thin-target simulations. In order to perform calculations for practical applications, which may involve complex geometries and multiple composite materials, nuclear models need to be embedded into a sophisticated transport code. The combination of nuclear models with a Monte Carlo transportation code like MCNP [5], GEANT4 [6,7] and FLUKA [8,9], nucleon meson transport codes (NMTC) [10] are widely utilized for designing facilities of engineering applications of the spallation reaction.

However, the nuclear models embedded into transportation codes need to be validated by experimental measurements. The several laboratories [11-14] worldwide have constructed specific experimental setups for measuring the double-differential cross sections and the spectra of neutrons produced in proton induced spallation reactions. In their experiments, a time-of-flight technique with organic liquid scintillators (or plastic scintillators) or/and a recoil proton measurement combined with a magnetic spectrometer has been utilized for neutron measurements. Those experimental measurements have made great contributions to the improvement of nuclear models. A Neutron Time-Of-Flight (NTOF) spectrometer, which is based on Heavy Ion Research Facility in Lanzhou (HIRFL), is designed for further investigations of reaction mechanisms and neutron productions of proton induced spallation reactions related to the ADS project.

In this work, the prediction ability of three nuclear models embedded in the GEANT4 for double-differential cross sections of the spallation neutron production is examined by comparing between the calculated results and the available experimental data [15]. The configuration and test beam results of the NTOF spectrometer are also presented and discussed in detail.

## II. Theoretical Calculations

GEANT4 is a Monte Carlo transport code developed in CERN for simulating the passage of particles through matter. It is widely used in particle and nuclear physics, accelerator physics, medical science, astrophysics and aerospace studies. In GEANT4, users have abundant choices of physics models to handle the interactions of particles with matter over an extended energy range.

Spallation reactions are usually described in two stages: intra-nuclear cascade and de-excitation. At the first stage, the incident particle transfers its kinetic energy to target nucleons by elastic collisions and a cascade of nucleon-nucleon collisions. Some of the particles that obtained enough energy to escape from the nucleus are emitted. The residual nucleus before the de-excitation stage can emit particles with the low energy, which are called pre-equilibrium particles. The energies of pre-equilibrium particles are generally greater than energies of particles emitted during the later stage. In the de-excitation stage, the excited nuclei loss their energy by evaporation of neutrons, protons or light charged particles. If the nuclei do not have enough energy to evaporate the particles, it may emit gamma rays and decay to the stable state.

In this work, the BIC [16], INCL [17] and QMD [18] models embedded in GEANT4 were used to calculate the first stage of the reactions. A statistical model is used for the second stage. The dynamical model is linked to the de-excitation handler provided by GEANT4. As a result, the double differential cross section of neutron production can be given at each stage. In Fig.1, the experimentally measured neutron production double-differential cross sections from the reaction on the Tungsten target induced by 800 MeV protons at detection angles of 30° to 150° are compared with simulated results of BIC, INCL and QMD models embedded in the GEANT4. The experimental data are taken from the EXFOR database [15]. Similar comparisons are given in Fig.2 for 800 MeV to 1600 MeV protons at 30° detection angle. The both of spectra multiply by a factor of $10^{-n}$ (n=0, 2, 4, 6) from top to bottom. The good agreements are achieved among the INCL calculations and experimental measurements in the all angles and energy range examined. The BIC and QMD calculations slightly underestimate the experimental results in some part of the neutron spectrum, especially in high energy tail. These discrepancies are related to the neutron detection angle and the incident proton energy.

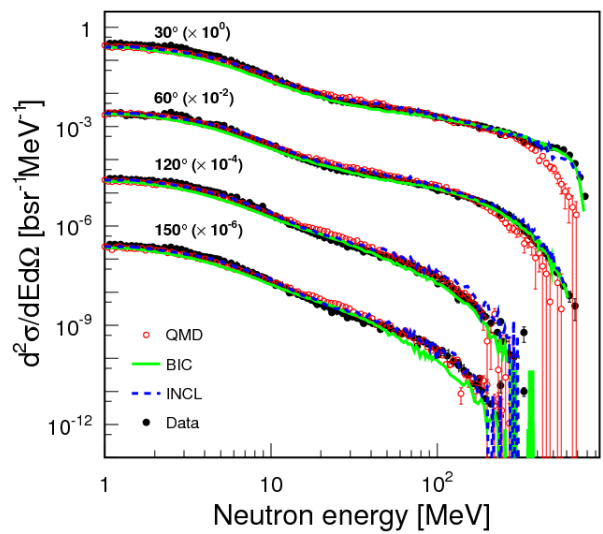

**Fig.1**. (Color online) Comparisons of experimental neutron production double-differential cross sections for 800 MeV proton on a Tungsten target with simulated results of different nuclear models at 30° to 150° detection angles

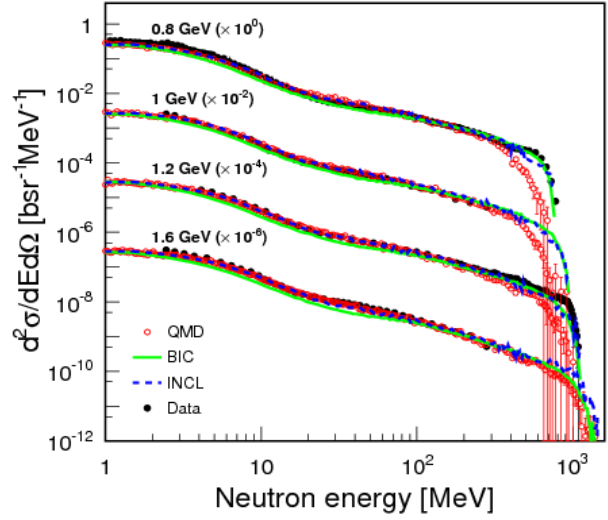

**Fig.2**. (Color online) Comparisons of experimental neutron production double-differential cross sections for 800 MeV to 1600 MeV protons on Tungsten target with simulated results of different nuclear models at 30° detection angle

## III. NTOF Spectrometer

The NTOF spectrometer is capable of measuring neutrons in a wide energy and angular ranges. It consists of a beam pick-up detector and 10 individual neutron detection modules. The schematic view of the NTOF spectrometer is shown in Fig.3. The beam pick-up detector is composed of BC404 plastic scintillator (5cm×5cm×0.2cm) detector with dual-PMT readout at both ends and is located upstream from the target. It is used to obtain the time and position information of the incident particle beam. Each of the neutron detection modules is composed of thin plastic scintillator detector (veto detector) and organic liquid scintillator detector. Two different sizes of cylindrical BC501A organic liquid scintillator detectors are placed in different angles. The larger BC501A detectors have 12.7cm in diameter and 12.7cm in length. They are used for detecting higher energy neutrons with longer neutron flight paths. The smaller BC501A detectors have a size of 5.08cm in diameter and 5.08cm in length, and they are used to measure low energy neutrons. In front of the individual BC501A detectors, BC404 plastic scintillator (15cm×15cm×0.3cm) coupled to 9813KB PMT are mounted as a veto detector to distinguish charged particle from non-charged particle (neutron and gamma-ray). For the separation of gamma-rays from neutrons, the pulse shape discrimination (PSD) property of organic liquid scintillator detector is utilized. Neutron kinetic energy is calculated from the time-of-flight (TOF) spectrum between the organic liquid scintillator detector and the beam pick-up detector. The neutron flight path from the target to neutron detector is about 1.5 m. The 400 ns TDC range is used in the experiment for measuring above 0.1 MeV neutrons.

A VME-based data acquisition system (DAQ) for recording the experimental data in event by event basis has been developed. The DAQ software is based on the CERN ROOT framework and running on Linux operating system. The main functions of the DAQ include data acquisition, online monitor and offline data analysis with a graphical user interface (GUI), which provides a convenient operation.

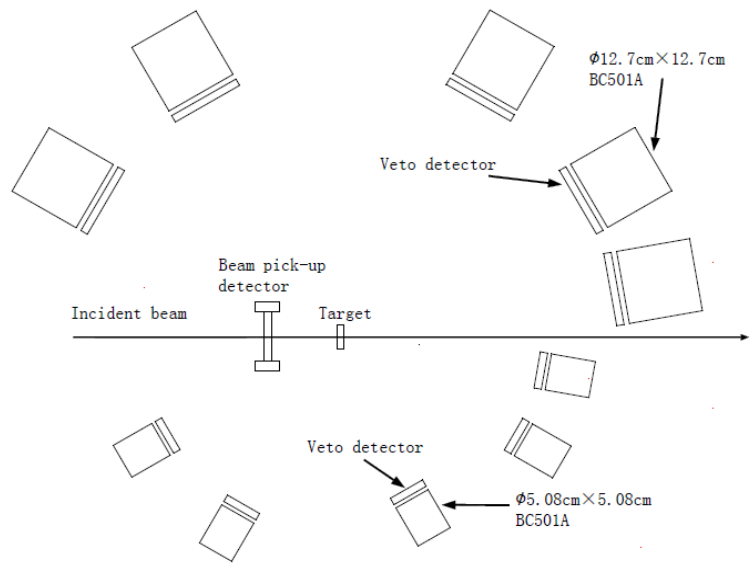

**Fig.3**. A schematic view of the NTOF spectrometer

IV. **Results and Discussion**

A test experiment for the whole system of the NTOF spectrometer was performed by measuring neutrons, gamma rays and charged particles in energy spectrum, production yield and angular distribution from a Tungsten target with $^{16}$O beam at a bombarding energy of 400 MeV/u.

In the experiment, the combination of light output spectra of the veto detector and BC501A scintillator detector was utilized for separation of charged particle events from non-charged particle (neutron and gamma-ray) ones, as shown in Fig. 4. The anode signal of BC501A scintillator detectors was divided into three pulses, and two of them were fed into charge-to-digital converters

(QDCs), which had different gate widths (total and slow). QDCs with the total and slow gates were used to eliminate the gamma-ray events with PSD by the two gate integration method. A typical result of PSD is shown in Fig. 5. The TOF spectra were measured between the beam pick-up detector and the BC501A detector. In Fig. 6, a typical TOF spectrum excluding the charged particle events is given. The prompt gamma peak of the TOF spectrum was used as a time reference of the spectrum. The energy calibration of organic liquid scintillator detectors was accurately determined by comparing the experimental light output of standard gamma sources with GEANT4 simulated one [19]. In Fig. 7, the comparisons of experimental neutron production yield at detection angles from 10° to 60° using larger BC501A detectors with GEANT4 results are shown. The neutron production yield is converted from TOF spectra with normalizing by unit solid angle and incident ion numbers. The experimental neutron spectrum shape are well reproduced the simulations. The experimental data have been normalized to the simulated one at 10°. The experimental results show that the whole system of the NTOF spectrometer works well, and the data analysis procedure is established.

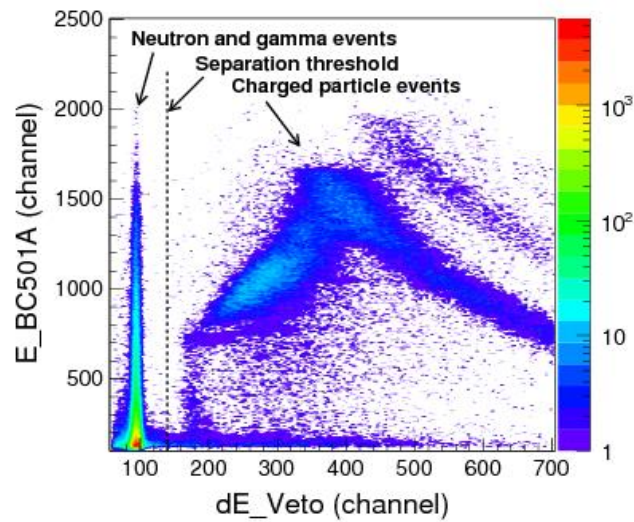

**Fig.4**. (Color online) Separation of charged particle events from non-charged particle (neutron and gamma-ray) ones, using the veto counter and a BC501A liquid scintillator

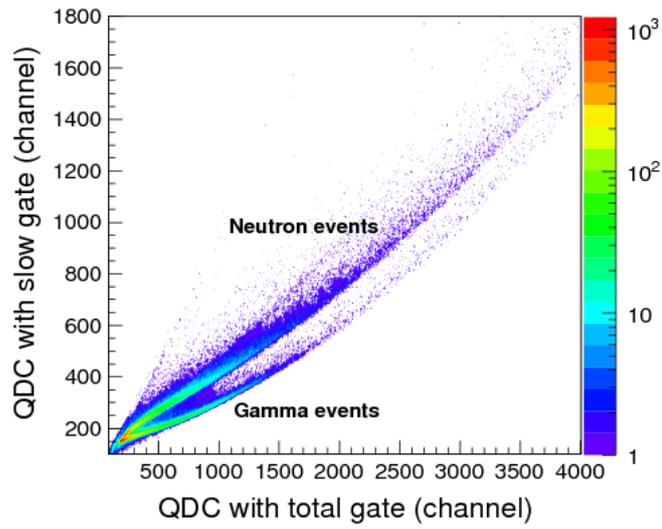

**Fig.5**. (Color online) A typical result of the pulse shape discrimination (PSD) in a BC501A liquid scintillator

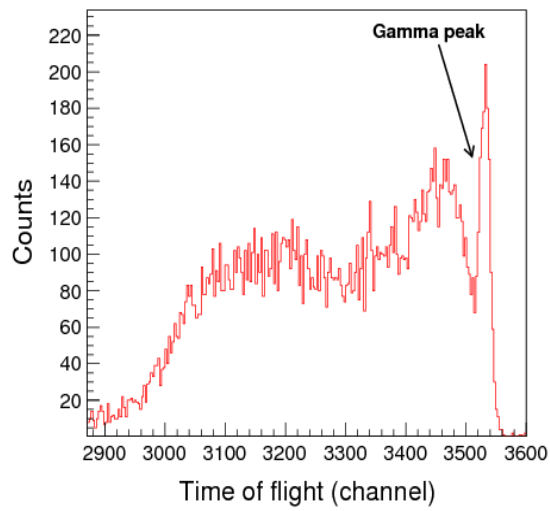

**Fig.6**. (Color online) Time-of-flight (TOF) spectrum for neutrons and gamma rays

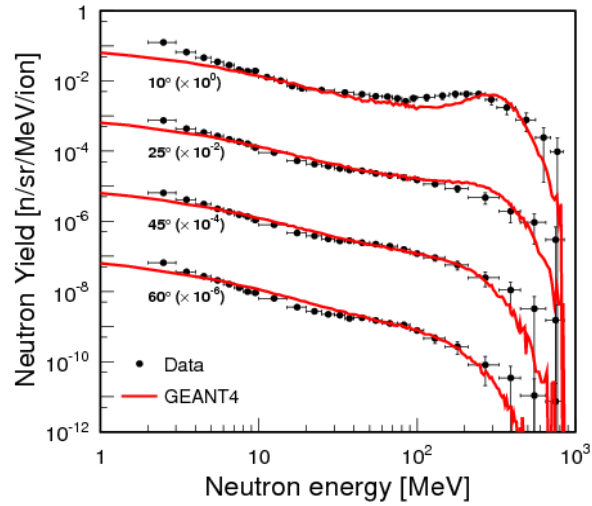

**Fig.7**. Comparisons of experimental neutron production yield with GEANT4 results for 400 MeV/u $^{16}$O bombarded on Tungsten at detection angles of 10° to 60°

V. **Conclusion**

For a design of nuclear engineering facilities, sophisticated simulation tools based on nuclear reaction models are required. In this work, the spallation neutron production cross sections of the BIC, INCL and QMD reaction models embedded in GEANT4 are compared with the available experimental data. The simulated results with INCL models are in better agreements with the experimental one in the entire energy and angular ranges studied. The predictions of the BIC and QMD models show slight discrepancies with the experimental results. In the design of the ADS project, accurate nuclear data are required. For this reason, the NTOF spectrometer was developed at Institute of Modern Physics, Chinese Academy of Sciences (IMP, CAS). A test run for the whole system of the spectrometer was performed by the experiment of 400 MeV/u $^{16}$O bombarded on Tungsten target. The experimental results show that the experimental apparatus works well, and the data analysis method is established. The further experiments will be done in future for the improvement and validation of nuclear models embedded in Monte Carlo transportation codes. These transportation codes will be the main simulation tools for the design of spallation target of China ADS project.